\documentclass[aps,prd,twocolumn,superscriptaddress,showpacs]{revtex4}

\usepackage{graphicx}
\usepackage{dcolumn}
\usepackage{bm}
\usepackage{color}
\usepackage{amsmath}
\usepackage{amstext}

\newcommand{\gsim}{ \mathop{}_{\textstyle \sim}^{\textstyle >} }
\newcommand{\lsim}{ \mathop{}_{\textstyle \sim}^{\textstyle <}}

\begin{document}

\title{Nuclear scattering of dark matter coupled to a new light scalar}

\author{Douglas P. Finkbeiner}
\email{dfinkbeiner@cfa.harvard.edu}
\affiliation{Harvard-Smithsonian Center for Astrophysics, 60 Garden St., Cambridge, MA 02138}

\author{Tracy R. Slatyer}
\email{tslatyer@fas.harvard.edu}
\affiliation{Physics Department, Harvard University, Cambridge, MA 02138}

\author{Neal Weiner}
\email{neal.weiner@nyu.edu}
\affiliation{Center for Cosmology and Particle Physics, Department of Physics, New York University, 
New York, NY 10003}

\date{\today}

\begin{abstract}

We consider the nuclear scattering cross section for the eXciting Dark Matter (XDM) model. In XDM, the Weakly Interacting Massive Particles (WIMPs) couple to the Standard Model only via an intermediate light scalar which mixes with the Higgs: this leads to a suppression in the nuclear scattering cross section relative to models in which the WIMPs couple to the Higgs directly. We estimate this suppression factor to be of order $10^{-5}$. The elastic nuclear scattering cross section for XDM can also be computed directly: we perform this computation for XDM coupled to the Higgs sector of the Standard Model and find a spin-independent cross section in the order of $4 \times 10^{-13}$ pb in the decoupling limit, which is not within the range of any near-term  direct detection experiments. However, if the XDM dark sector is instead coupled to a two-Higgs-doublet model, the spin-independent nuclear scattering cross section can be enhanced by up to four orders of magnitude for large $\tan \beta$, which should be observable in the upcoming SuperCDMS and ton-scale xenon experiments.

\end{abstract}

\pacs{95.35.+d}

\maketitle


\section{Introduction}

Motivated by the apparent excess of $e^+e^-$ annihilation in the
Galactic center \cite{Weidenspointner:2006nu, Weidenspointner:2007},
Finkbeiner \& Weiner \cite{Finkbeiner:2007kk} proposed a model of
eXciting Dark Matter (XDM) in which Weakly Interacting Massive
Particles (WIMPs) collisionally excite and subsequently de-excite via
$e^+e^-$ emission.  This model uses the kinetic energy of the WIMP
dark matter to create $e^+e^-$ pairs, in contrast with light dark
matter models in which the pairs result from the mass energy of WIMP
annihilation \cite[e.g.,][]{Boehm:2003bt}, where the WIMP mass must be
less than a few MeV \cite{Beacom:2004pe, Beacom:2005qv}.  Because the
XDM WIMP must have a weak-scale mass ($\sim 500$ GeV) it retains many
of the desirable properties of weak-scale WIMPs such as the thermal
relic freeze-out abundance.  Collisional excitation of weak-scale
WIMPs was also discussed by Pospelov \& Ritz \cite{Pospelov:2007xh}.

The recent claim by DAMA/LIBRA of an annual modulation signal \cite{2008arXiv0804.2741B}
conflicts strongly with other direct detection experiments [CDMS, XENON-10,
ZEPLIN-II, CRESST], under the assumption of elastic nuclear scattering.  Models in
which the WIMP scatters \emph{inelastically}, exciting an internal state of
order 100 keV, appear to accommodate all available data  \cite{2008arXiv0807.2250C}.  The success of
inelastic WIMP scattering in these two cases (albeit with different mass
splittings) raises the question of whether a nuclear scattering cross section
of $10^{-4}$ pb (required by Chang et al \cite{2008arXiv0807.2250C}) could arise naturally in the XDM model.

In XDM, WIMPs couple to the Standard Model via a new light scalar which mixes with the Higgs. Preliminary results from ATIC \cite{ATIC2005} and PAMELA motivate the existence of just such a light particle \cite{sommerfeldpaper}. However, the mixing angle between the light scalar and the Higgs suppresses the nuclear scattering cross section relative to generic models of dark matter where the Higgs couples directly to the WIMPs. We first estimate this suppression factor, then calculate the nuclear scattering cross section for the XDM dark sector coupled to the Standard Model or a two-Higgs-doublet model, and finally discuss the role of inelastic scattering and prospects for direct detection.

\section{Summary of the XDM model}
The defining feature of the XDM model is that the WIMP has an excited
state which can be collisionally excited, and subsequently decay to produce
$e^+e^-$ pairs.  The excited state could exist due to compositeness of
the dark matter, or arise from an approximate symmetry of the theory.

For the excited state to be accessible in the Milky Way, and relevant
for $e^+e^-$ production, only a narrow kinematical range must be
considered for the mass splitting, $\delta$.  For the decay to the
ground state to be energetically capable of producing $e^+e^-$ pairs,
one must have $\delta > 1.022$ MeV. On the other hand, the kinetic
energy available for a pair of $500$ GeV WIMPs colliding each with
velocities $v \sim 600$ km/s (roughly the escape velocity of the
galaxy), is $2$ MeV, setting an upper bound on $\delta$.

To produce a sufficiently high number of positrons to explain the
INTEGRAL signal, a large cross section is required
\cite{Finkbeiner:2007kk,Pospelov:2007xh}, comparable to the geometric
cross section set by the characteristic momentum transfer. That is,
$\sigma \sim (M_\chi \delta)^{-1}$ is of the correct size. Such a
cross section can arise naturally \cite{Finkbeiner:2007kk}, but
requires the presence of a new light scalar $\phi$, with $m^2_\phi
\lsim M_\chi \delta$. The $\chi$ can excite by emitting a $\phi$ with
amplitude $\lambda_-$ or can scatter elastically with amplitude
$\lambda_+$. We generally assume $\lambda_-\sim \lambda_+$, but this
need not necessarily be the case.

The XDM dark-sector Lagrangian takes the form \cite{Finkbeiner:2007kk},
\begin{eqnarray} \mathcal{L}_{\text{XDM}} & = & \frac{1}{2} \partial_{\mu} \phi \partial^{\mu} \phi + \chi_i^{\dagger} \sigma_{\mu} \partial^{\mu} \chi_i - m_D \chi_1 
\chi_2 - \nonumber \\ & & - \lambda_1 \phi \chi_1^2 -  \lambda_2 \phi \chi_2^2 - V(\phi) + \alpha_{\text{XDM}} \phi^2 h^2. \label{XDM_dark_lagrangian} \end{eqnarray}
The last term couples the dark sector to the Standard Model via the Higgs. Defining new basis states $\chi, \chi^* = 1/\sqrt{2} (\chi_1 \mp \chi_2)$, and new couplings $\lambda_\pm = \lambda_1 \pm \lambda_2$, the Lagrangian becomes
\begin{eqnarray} \mathcal{L}_{\text{XDM}} & = & \frac{1}{2} \partial_{\mu} \phi \partial^{\mu} \phi + \chi^{\dagger} \sigma_{\mu} \partial^{\mu} \chi + \chi_*^{\dagger} \sigma_{\mu} \partial^{\mu} \chi_* - \nonumber \\ & & - \frac{1}{2} \left(- m_D \chi^2 + m_D \chi_*^2 + \lambda_+ \phi \chi^2 + \lambda_+ \phi \chi_*^2 \right) - \nonumber \\ & & - \lambda_- \phi \chi \chi_* - V(\phi) + \alpha_{\text{XDM}} \phi^2 h^2. \label{XDM_lagrangian} \end{eqnarray}
We write $\alpha_{\text{XDM}} = m_{\phi}^2 K / 2 v_h^2$, where $K$ is the tuning parameter that determines the correction to the $\phi$ VEV and $v_h$ is the VEV of the Higgs $h$ ($v_h = 246$ GeV in the Standard Model): naturalness demands that $K \le 100$ (corresponding to a 1\% fine-tuning), and $\alpha_{\text{XDM}} \le 10^{-3}$. The mass splitting $\delta$ is given by $\delta = 2 \lambda_+ \langle \phi \rangle$, where $\langle \phi \rangle$ is the vacuum expectation value of the $\phi$ scalar. While XDM in its current form is an ad hoc extension to the Standard Model and is not supersymmetric, we expect it to be possible to realize the XDM mechanism in supersymmetric models as well.

\section{Effect of the Intermediate Light Scalar on the Nuclear Scattering Cross Section}

\begin{figure}[h]
\centering
\includegraphics[width=0.3\textwidth]{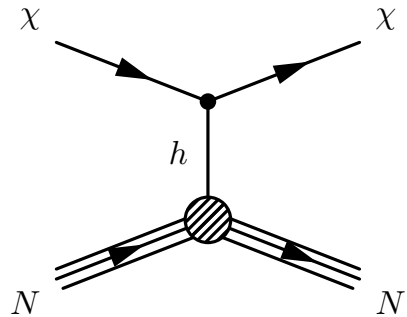}
\caption{t-channel WIMP-nucleus scattering via Higgs transfer.}
\label{t-higgs}
\end{figure}

XDM WIMPs couple to the visible sector by the mixing of the Higgs with the new light scalar $\phi$, so at tree level, WIMP-nucleus scattering occurs only via $t$-channel $\phi$-Higgs exchange. A similar scenario has been explored in \cite{Barger:2007im}, where the dark matter is itself a scalar that couples to the Higgs. In generic models of dark matter other scattering channels are available (such as $t$-channel $Z$ exchange, and $s$ and $t$-channel squark exchange in the Minimal Supersymmetric Standard Model (MSSM)), but the $t$-channel Higgs exchange (Fig. \ref{t-higgs}) generally provides the dominant contribution \cite{1993PhRvD..48.3483D}. Consequently, the WIMP-nucleus scattering cross section in XDM is suppressed mostly by the mixing angle between the new light scalar and the Higgs, and the suppression factor can be calculated in a straightforward way.

\begin{figure}[h]
\centering
\includegraphics[width=0.3\textwidth]{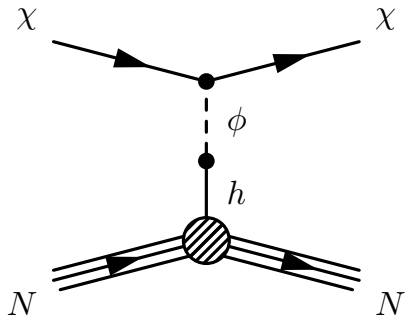}
\caption{t-channel WIMP-nucleus scattering via $\phi$-mediated Higgs transfer (XDM scenario).}
\label{t-higgs-XDM}
\end{figure}

The Feynman diagram for $t$-channel Higgs-mediated WIMP-nucleus scattering in XDM (Figure \ref{t-higgs-XDM}) is identical to Figure \ref{t-higgs} except that the $\chi$ and $\chi_*$ WIMPs do not couple directly to the Higgs: their interaction with the Higgs is mediated by the relatively light scalar $\phi$ boson, which then mixes with the Higgs. Integrating out the $\phi$ propagator gives rise to an effective WIMP-Higgs interaction with amplitude $\lambda_+ . 4 \alpha_{\text{XDM}} v_h \langle \phi \rangle / (m_{\phi}^2 - t)$. The analysis is identical for elastic scattering of $\chi^*$ particles; for the inelastic de-excitation $\chi_* A \rightarrow \chi A$ (where $A$ denotes the nucleus) the only change is that $\lambda_+$ is replaced with $\lambda_-$. (The inelastic excitation $\chi A \rightarrow \chi_* A$ is kinematically forbidden for virtually all the WIMPs in the Earth's neighborhood.) In the center-of-mass frame $t = - 2 |\mathbf{p}|^2 (1 - \cos \theta)$, where $\theta$ is the scattering angle and the incoming WIMP has 4-momentum $\mathbf{p}$. For the energy range of interest, the momentum transfer is generally small compared to $m_{\phi}$, so the effective coupling becomes,
\begin{equation} g_{h \chi \chi} = 4 \lambda_+ \alpha_{\text{XDM}} v_h \langle \phi \rangle / m_{\phi}^2 = \frac{\delta K}{v_h}. \end{equation}
(If $m_{\phi}$ is very small, of the same order as the momentum transfer, then the $\phi$ exchange can introduce a simple angular dependence to the cross section.)

This effective coupling is very small: allowing $K$ to take values up to $100$, we find,
\begin{equation} g_{h \chi \chi} \approx 4.1 \times 10^{-4} \left( \frac{K}{100} \right) \left(\frac{\delta}{1 \text{MeV}}\right) \left( \frac{v_h}{246 \text{GeV}} \right)^{-1}. \end{equation}
Suppose we compare XDM to a generic model with $g_{h \chi \chi} = 0.1$ as a benchmark. Then we see the XDM nuclear scattering cross section (proportional to the square of the coupling) is suppressed by at least five orders of magnitude.

\section{Calculation of the XDM WIMP-nucleus elastic scattering cross section}

We now calculate the XDM WIMP-nucleus
cross section directly, for XDM coupled to the Higgs sector of the Standard Model. We will compute
the effective spin-independent cross section for low-energy scattering between a
WIMP and a pointlike nucleus. (To determine the actual
scattering cross section this result must then be integrated over the nuclear form factor.) 

Evaluating the elastic scattering amplitude $\mathcal{M}$ in the low-energy ($ t \ll
m_{\phi} \ll m_{h}$)
nonrelativistic regime, for a nucleus with $Z$ protons and $A-Z$ neutrons, we obtain,
\begin{equation} |\mathcal{M}| = (2 M)(2 m_{\chi}) \frac{g_{h \chi \chi}}{m_h^2} \left( Z g_{hpp} + (A-Z) g_{hnn} \right),
\end{equation}
where $g_{hpp}$ and $g_{hnn}$ are the couplings of the Higgs to the
proton and neutron respectively, and $M$ is the nuclear mass. Thus the cross section due to Higgs exchange is given by,
\begin{equation} \sigma = \frac{\mu_{A \chi}^2}{\pi} \left( \frac{K \delta}{v_h} \right)^2 \left( \frac{Z g_{hpp} + (A-Z) g_{hnn}}{m_h^2} \right)^2, \end{equation}
where $\mu_{A \chi}$ is the reduced mass of the system.

The couplings of the Higgs to the proton and neutron are given by \cite{2004JCAP...07..008G, 1993PhRvD..48.3483D},
\begin{eqnarray} g_{hNN} & = & \sum_{q = u, d, s, c, b, t} \langle N |\bar{q} q|
N \rangle g_{hqq} \\  \langle N |\bar{q} q|
N \rangle & = & f^N_{Tq} m_N / m_q, \end{eqnarray}
where $N = p, n$, and the nucleon parameters $f^N_{Tq}$ are given by \cite{1996PhR...267..195J, Cheng:1988im, 1991PhLB..253..252G} 
\begin{equation} f^p_{Tu} =  0.023, \quad  f^p_{Td}  = 0.034, \quad
 f^p_{Ts}  = 0.14, \nonumber \end{equation}
 \begin{equation}  f^p_{Tc}  =  f^p_{Tb}  =  f^p_{Tt}  =
 0.0595, \end{equation}
 \begin{equation}  f^n_{Tu} =  0.019, \quad  f^n_{Td}  = 0.041, \quad
 f^n_{Ts}  = 0.14, \nonumber \end{equation}
 \begin{equation} f^n_{Tc}  =  f^n_{Tb}  =  f^n_{Tt}  =
 0.0592. \end{equation}
 Note that $f^N_{Ts}$ is not well determined \cite{1996PhR...267..195J}: values in the literature also include 0.08 \cite{1992NuPhB.387..715H} and 0.46 \cite{1988PhRvD..38.2869C, Cheng:1988im}. We adopt the value obtained by Gasser et al \cite{1991PhLB..253..252G} and used in the DarkSUSY package \cite{2004JCAP...07..008G}. The nucleon parameters for the heavy quarks c, b, t are determined by those for the light quarks \cite{1978PhLB...78..443S,1996PhR...267..195J}, so the uncertainty in $f^N_{Ts}$ carries over to those parameters as well. However, choosing different values for $f^N_{Ts}$ modifies the final cross section by a factor $\sim 3-4$ at most.
 
In the Standard Model, the quark-Higgs Yukawa couplings are given by $g_{hqq} = - m_q / v_h$, so the nuclear scattering cross section takes the form,
\begin{equation} \sigma = \frac{K^2}{\pi} \left( \frac{\mu_{A \chi} \delta}{v_h^2 m_h^2} \right)^2 \left(0.3776 (A - Z) m_n + 0.3755 Z m_p \right)^2. \end{equation}

To compare with the current limits from direct detection experiments, we now take $A=Z=1$ and compute the scattering cross section for a single proton. In this case $m_p = M \ll m_{\chi}$ so $\mu_{A \chi} \approx m_p$. The lab-frame low-energy elastic scattering cross section is then,
\begin{eqnarray} \sigma & \approx & 4 \times 10^{-13} \text{pb} \times \left(
\frac{m_h}{100 \text{GeV}} \right)^{-4} \times \left( \frac{K}{100} \right)^2 \times \nonumber \\ & & \times \left( \frac{\delta}{1 \text{MeV}} \right)^2 \times \left( \frac{v_h}{246 \text{GeV}} \right)^{-4}. \label{XDM-sigma}\end{eqnarray}
This is a very small cross section, some five orders of magnitude below current experimental limits.

\section{Two-Higgs-Doublet Models}

If the XDM dark sector is instead coupled to a two-Higgs-doublet model (2HDM), the cross section can potentially be significantly enhanced by interactions involving the heavy Higgs. Such models constitute one of the simplest extensions of the Standard Model that allow additional CP violation \cite{PhysRevD.8.1226,PhysRevLett.37.657}, and have been discussed extensively in the literature (see \cite{Accomando:2006ga} for a review). We will concern ourselves here with type-II 2HDMs, where one Higgs doublet couples purely to up-type fermions and the other couples purely to down-type fermions \cite{PhysRevD.19.945}: the MSSM is one example of such a model. The effective coupling between $\chi$ and the heavy Higgs can be quite large for large $\tan{\beta}$ in such models, as the smaller VEV results in a weaker constraint on the $\phi$-Higgs mixing $\alpha_{\text{XDM}}$. 

Such a model requires the Lagrangian (Eq. \ref{XDM_lagrangian}) to be modified to take couplings to the second Higgs into account. For example, let us replace the usual $\phi$-Higgs mixing term with $\alpha_{\text{XDM}} \phi^2 H^0_u H^0_d + h.c.$, where as usual the VEVs of the neutral Higgses are defined by,
\begin{equation} \langle H^0_u \rangle = \frac{v_u}{\sqrt{2}}, \qquad \langle H^0_d \rangle  =  \frac{v_d}{\sqrt{2}}, \end{equation} 
and $\tan{\beta} = v_u / v_d$, $v_h^2 = v_u^2 + v_d^2 \approx (246 \text{GeV})^2$. The $H^0_u$-$\phi$ and $H^0_d$-$\phi$ mixing terms then become, respectively, $(2 \alpha_{\text{XDM}} \langle \phi \rangle v_d/\sqrt{2}) \phi (H^0_u + h.c.)$ and $(2 \alpha_{\text{XDM}} \langle \phi \rangle v_u/\sqrt{2}) \phi (H^0_d + h.c.)$. The constraint on $\alpha_{\text{XDM}}$ now comes from the term $2 (\alpha_{\text{XDM}} v_u v_d/2) \phi^2$, and can be written as,
\begin{equation} \alpha_{\text{XDM}} = K \frac{m_{\phi}^2}{2 v_u v_d}, \end{equation}
where as previously $K \le 100$. Thus the Higgs-$\phi$ mixing terms in the Lagrangian can be written as,
\begin{equation} \mathcal{L}_{\text{mixing}} = \frac{1}{\sqrt{2}} K m_{\phi}^2 \langle \phi \rangle \times \left( \frac{1}{v_u} H^0_u + \frac{1}{v_d} H^0_d \right) \phi + h.c. \end{equation}

We are interested in scatterings mediated by the scalar uncharged physical Higgs bosons. We now transform to the basis of mass eigenstates, $h^0$ and $H^0$, which as usual are related to $H^0_u$ and $H^0_d$ by a mixing angle $\alpha$,
\begin{eqnarray} \Re{H^0_u} & = & \frac{1}{\sqrt{2}} \left( \cos{\alpha} \, h^0 + \sin{\alpha} \, H^0 \right), \nonumber \\  \Re{H^0_d} & = & \frac{1}{\sqrt{2}} \left( - \sin{\alpha} \, h^0 + \cos{\alpha} \, H^0 \right). \end{eqnarray}

Thus the mixing terms become,
\begin{eqnarray} \mathcal{L}_{\text{mixing}} & = & K m_{\phi}^2 \langle \phi \rangle \times \left( \left( \frac{\cos{\alpha}}{v_u} - \frac{\sin{\alpha}}{v_d} \right) h^0 + \right. \nonumber \\ & & + \left. \left( \frac{\sin{\alpha}}{v_u} + \frac{\cos{\alpha}}{v_d} \right) H^0 \right) \phi. \end{eqnarray}

The effective Higgs-WIMP interactions can then be written,
\begin{eqnarray} g_{h^0 \chi \chi} & = & \frac{K \delta}{2} \left( \frac{\cos{\alpha}}{v_u} - \frac{\sin{\alpha}}{v_d} \right) \nonumber \\ g_{H^0 \chi \chi} & = & \frac{K \delta}{2} \left( \frac{\sin{\alpha}}{v_u} + \frac{\cos{\alpha}}{v_d} \right). \end{eqnarray}

The quark-Higgs Yukawa couplings $g_{hqq}$ for the light CP-even Higgs are given by \cite{1996PhR...267..195J},
\begin{equation} g_{hqq} = \frac{m_q}{v_h} \times \left \{ \begin{array}{ll} - \frac{\cos{\alpha}}{
\sin{\beta}} & \text{for $q=$ u, c, t}, \\ \frac{\sin{\alpha}}{\cos{\beta}} &
\text{for $q=$ d, s, b}, \end{array} \right. \end{equation}
whereas those for the heavy CP-even Higgs are,
\begin{equation} g_{Hqq} = \frac{m_q}{v_h} \times \left \{ \begin{array}{ll} - \frac{\sin{\alpha}}{
\sin{\beta}} & \text{for $q=$ u, c, t}, \\ - \frac{\cos{\alpha}}{\cos{\beta}} &
\text{for $q=$ d, s, b}. \end{array} \right. \end{equation}
The Standard Model couplings are recovered in the decoupling limit where $\alpha \approx \beta - \pi/2$.

The effective WIMP-nucleon interaction for a proton thus becomes,
\begin{eqnarray} g_{p \chi \chi} & = & \frac{K \delta m_{p}}{2 v_h^2} \left( \frac{1}{m_{h^0}^2} \left( \frac{\cos{\alpha}}{\sin{\beta}} - \frac{\sin{\alpha}}{\cos{\beta}} \right) \times \right. \nonumber \\ & & \times \left(0.2335 \frac{\sin{\alpha}}{\cos{\beta}} - 0.142 \frac{\cos{\alpha}}{\sin{\beta}} \right) + \nonumber \\ & & + \frac{1}{m_{H^0}^2} \left( \frac{\sin{\alpha}}{\sin{\beta}} + \frac{\cos{\alpha}}{\cos{\beta}} \right) \times \nonumber \\ & & \left. \times \left(- 0.142 \frac{\sin{\alpha}}{\sin{\beta}} -
0.2335 \frac{\cos{\alpha}}{\cos{\beta}} \right) \right) \label{two-higgs-coupling} \end{eqnarray}
The nuclear scattering cross section normalized to a single nucleon is, as previously,
\begin{equation} \sigma = \frac{\mu_{A \chi}^2}{\pi} g_{p \chi \chi}^2. \end{equation}

At large $\tan \beta$, Eq. \ref{two-higgs-coupling} can give rise to a much greater cross section than in the Standard Model case (Eq. \ref{XDM-sigma}), via the contribution from the heavy Higgs. In the most general two-Higgs-doublet model, the Higgs masses and mixing angles are all independent parameters: however, allowing all these parameters to vary freely can lead to conflicts with experimental constraints, and problematic flavor changing neutral current processes. To estimate the effect of increasing $\tan \beta$, we consider the tree-level MSSM relations between $m_H, m_h, \alpha$ and $\beta$ \cite{1996PhR...267..195J},
\begin{eqnarray} m_{h,H}^2 & = & \frac{1}{2} \left(m_A^2 + M_Z^2 \mp \right. \nonumber \\ & & \left. \mp \sqrt{ \left(m_A^2 + M_Z^2 \right)^2 - 4 M_Z^2 m_A^2 \cos^2{2 \beta}} \right), \end{eqnarray}
\begin{equation} \frac{\sin{2 \alpha}}{\sin{2 \beta}} = - \frac{m_A^2 + m_Z^2}{m_H^2 - m_h^2}, \quad \frac{\cos{2 \alpha}}{\cos{2 \beta}} = - \frac{m_A^2 - m_Z^2}{m_H^2 - m_h^2}. \end{equation}
Note that these relations are only approximately correct: the Higgs masses experience significant radiative corrections. 

\begin{figure}
\centering
\includegraphics[width=0.49\textwidth]{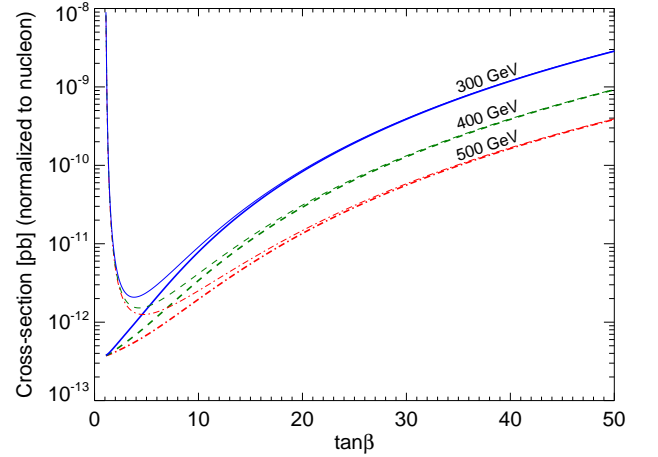}
\caption{Variation of the cross section with $\tan \beta$ for XDM coupled to a two-Higgs-doublet model, for $m_A = 300, 400, 500$ GeV. The thick lines correspond to cases where the light Higgs mass is held constant at $100$ GeV, whereas the thin lines use the tree-level MSSM values.}
\label{twohiggsfig}
\end{figure}

Fig. \ref{twohiggsfig} shows the variation of the spin-independent cross section as a function of $\tan \beta$, for different pseudoscalar Higgs masses $m_A$ (see \cite{Eriksson:2008cx} for a recent analysis of constraints on $m_A$). We also show the cross section where the light Higgs mass is held constant at $100$ GeV (but the tree-level relations are used to obtain the mixing angle $\alpha$), to facilitate comparison with the Standard Model case discussed previously and to explore the effect on the nuclear scattering cross section of varying the light Higgs mass. At large $\tan \beta$ we find that varying the mass of the light Higgs has little impact, as the contribution from exchange of the heavy Higgs dominates.

\begin{figure}[ht]
\centering
\includegraphics[width=0.4\textwidth]{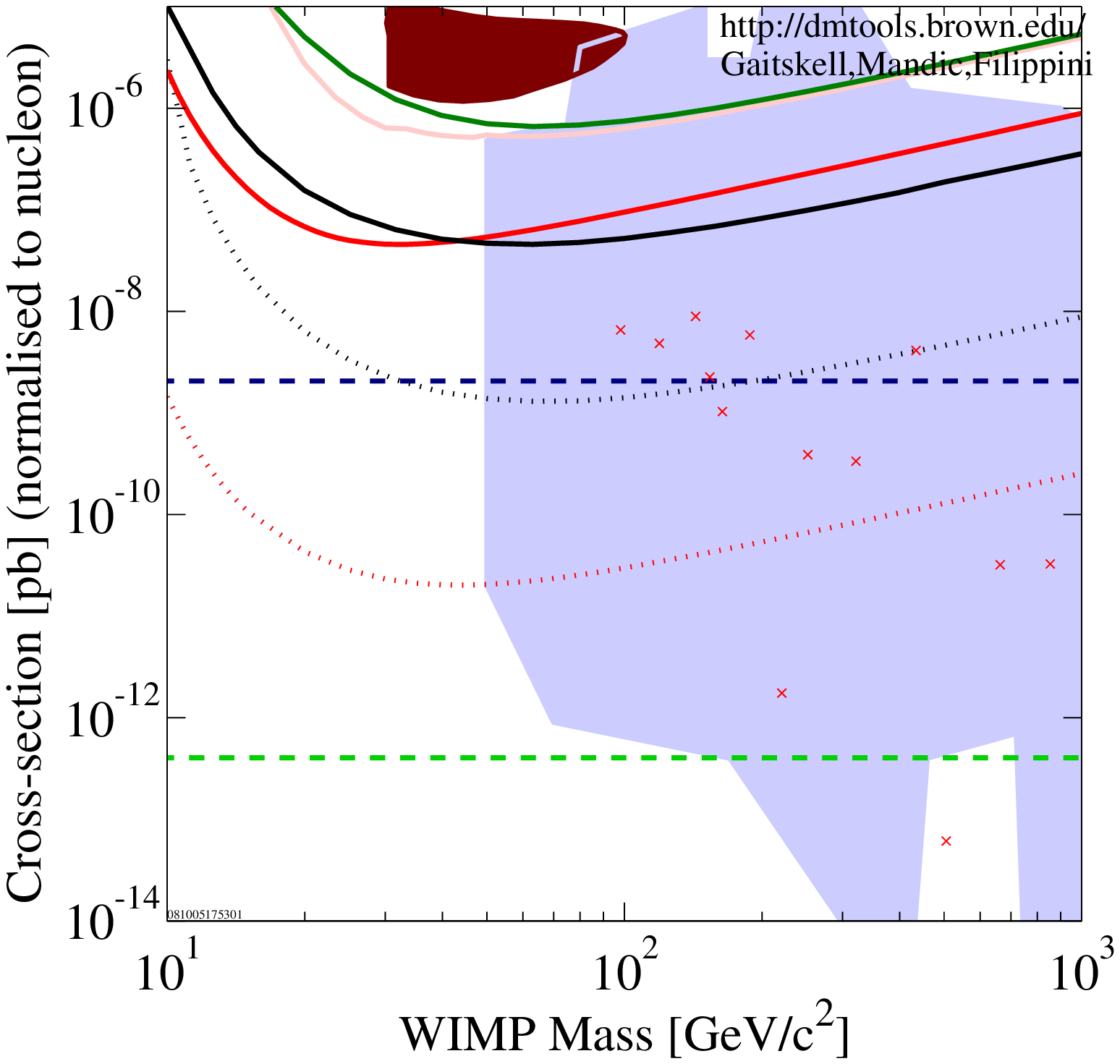}
\includegraphics[width=0.4\textwidth]{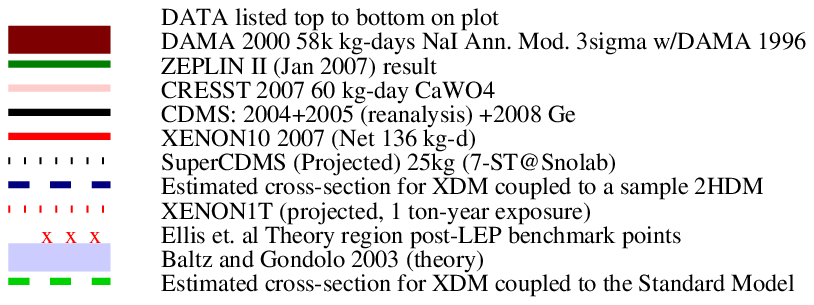}
\caption{Current experimental limits and theoretical predictions for the spin-independent proton-WIMP elastic scattering cross section. The lower dashed line is the estimated cross section for XDM coupled to the Standard Model, calculated in this paper. The upper dashed line is the cross section for XDM coupled to a 2HDM with $\tan \beta = 50$ and $m_A = 300$ GeV, as shown in Fig. \ref{twohiggsfig}.}
\label{detection-space}
\end{figure}

\section{Inelastic Scattering}

Tucker-Smith and Weiner \cite{2008arXiv0807.2250C, Smith:2001hy, Tucker-Smith:2004jv} have investigated models of inelastically scattering dark matter in which the modulation observed by DAMA/LIBRA \cite{2008arXiv0804.2741B} remains consistent with the absence of signals in other direct detection experiments, due to kinematic suppression of the scattering rate. XDM does not satisfy the requirements for those models, as the mass splitting $\delta$ is too large by a factor of 10: nonetheless, it is interesting to consider how inelastic scattering processes might contribute to the interaction cross section.

The coupling constant for inelastic scattering in XDM, $\lambda_-$, may be significantly larger than $\lambda_+$ \cite{2008arXiv0805.3531F}, so the low cross section calculated for elastic scattering does not necessarily limit the inelastic scattering cross section. The cross section for dark matter inelastically scattering with itself ($\chi \chi \rightarrow \chi \chi^*$) can be quite large in XDM: however, this is mostly due to enhancement by ladder diagrams \cite{Finkbeiner:2007kk}. Such an enhancement does not apply to WIMP-nucleus scattering, as each $\phi$-$h$ exchange is suppressed by a factor of $\alpha_{\text{XDM}}$. 

The excitation process $\chi A \rightarrow \chi^* A$ is suppressed by a kinematic factor $\sqrt{1 - 2 \delta / \left( \mu_{A
\chi} v^2 \right) }$, where $v$ is the lab-frame velocity of the incident WIMP, and the cross section falls to zero at the threshold velocity $v = \sqrt{2 \delta / \mu_{A \chi}}$ (in units of $c$): for a mass splitting of $\delta = 1$ MeV and any reasonable choice of nucleus, the threshold velocity is higher than the velocity of all but a minute fraction of the WIMPs in the neighborhood of the Earth. While the de-excitation process is not kinematically suppressed, the density of ambient $\chi^*$ particles is extremely low compared to that of the $\chi$ particles \cite{2008arXiv0805.3531F}, so the contribution to the detection rate is negligible even if the cross section is large.

Clearly, XDM in its current form does not predict significant inelastic nucleus-WIMP scattering, due to the relatively large mass splitting. However, a modified XDM model containing multiple excited states, with some splittings of order $100$ keV, might permit a greater degree of inelastic scattering, and support the Tucker-Smith-Weiner mechanism (such models have been discussed in \cite{sommerfeldpaper,nimaneal}). The strong suppression of the elastic scattering cross section calculated in this work may be useful in such a model: if inelastic scattering effects are to reconcile DAMA with other direct detection experiments, the elastic scattering cross section must be very small relative to inelastic scattering contributions.

\section{Prospects for Direct Detection}

Figure~\ref{detection-space} \cite{DMlimits} shows the current and projected upper limits on the elastic scattering cross section (normalized to a single nucleon) for an array of experiments as a function of the WIMP mass, as well as the predicted values obtained by sampling the parameter spaces of theoretical models of dark matter. It is clear that the prediction for XDM coupled to the Standard Model falls well below all current constraints, and also well below the projected lower limits of the SuperCDMS and ton-scale xenon experiments. However, if the XDM dark sector is instead coupled to a two-Higgs-doublet model with large $\tan \beta$ ($\gsim 20-30$), the cross section may be large enough to permit direct detection in upcoming experiments.

\section{Conclusion}

The XDM model, where the dark sector couples to the Standard Model only via the weak mixing of a light scalar with the Higgs, has a highly suppressed cross section for elastic nuclear scattering relative to models where the WIMP couples directly to the Higgs. We estimate this suppression at five or more orders of magnitude. Computing the XDM nuclear scattering cross section directly, for XDM coupled to the Standard Model, leads to a WIMP-proton cross section for elastic scattering of, 
\begin{eqnarray} \sigma & \approx & 4 \times 10^{-13} \text{pb} \times \left(
\frac{m_h}{100 \text{GeV}} \right)^{-4} \times \left( \frac{K}{100} \right)^2 \times \nonumber \\ & & \times \left( \frac{\delta}{1 \text{MeV}} \right)^2 \times \left( \frac{v_h}{246 \text{GeV}} \right)^{-4}, \end{eqnarray} 
which is beyond the reach of future proposed direct detection experiments. However, coupling the XDM dark sector to a two-Higgs-doublet model can give rise to interesting cross sections, provided $\tan \beta$ is large.

Models of dark matter incorporating inelastic scattering may be of interest in reconciling the results of DAMA/LIBRA with other direct detection experiments, but XDM in its current form is unsuitable, as the mass splitting is too large for the excited state to be kinematically accessible by WIMPs in the neighborhood of the Earth. However, extensions of the model containing multiple nearly degenerate excited states might allow inelastic scattering to play a greater role.

\onecolumngrid
\bibliography{xdm_nuc_scat}
\bibliographystyle{apsrev}

\end{document}